\documentclass[12pt]{article}
\begin{document}
\hfill
{\vbox{
\hbox{UCI-TR-96-32}
\hbox{UTEXAS-HEP-96-12}
\hbox{DOE-ER-40757-081}}}

\begin{center}
{\Large \bf Quark Anomalous Chromomagnetic Moment Bounds - Projection to
Higher Luminosities and Energy}
\footnote{
Submitted to the Proceedings of the 1996 DPF/DPB Summer Study on New 
Direction for High Energy Physics, Snowmass CO.
Work supported in part by the U. S. Department of Energy under Contracts
No. DE-FG03-93ER40757 (K.C.) and No. DE-FG03-91ER40679 (D.S.).
}

\vspace{0.15in}

Kingman Cheung\\ 

{\it Center for Particle Physics, University of Texas, Austin, Texas 78712} \\

Dennis Silverman\\ 

{\it Dept. of Physics and Astronomy, U.C. Irvine, Irvine, CA 92697-4575}

\end{center}
\thispagestyle{empty}

\begin{abstract} 
The statistical limits on detectability of an anomalous 
chromomagnetic moment of a quark coupling to a gluon are projected
to higher luminosities at the Tevatron at Fermilab, and to the LHC.
They roughly scale as the energy, and are not strongly improved with
increasing lumonisity.
\end{abstract}

\section{Anomalous Chromomagnetic Moments of Quarks}

New interactions or composite structure can lead to anomalous magnetic
moments for electromagnetic interactions and anomalous chromomagnetic
moments for colored intermediate states of colored quarks.  The form of
the interaction is
\begin{equation}
{\cal L}_{eff} = g \bar{\psi} \frac{\lambda_a}{2}
(-\gamma_\mu G_\mu^a + \frac{\kappa'}{2} \sigma_{\mu \nu} G_{\mu \nu}^a)
\psi
\end{equation} 
It has been shown\cite{silverman} that these could account for a
possible discrepancy between CDF and D0 data and QCD\cite{CDF,tung},
although this can also be accounted for by larger gluon structure
functions.

Since these same interactions can contribute to the mass of the
quarks, they are usually considered to be small for light mass
quarks\cite{brodskydrell,shawsilvermanslansky}, but can be larger for
heavier quarks, such as the $b$ and certainly for the $t$ quark.  New
heavy mass intermediate fermions could be allowed, however, if
balanced by much heavier bosons since the ratio $\kappa' \propto
m_F/m_B^2$ occurs\cite{brodskydrell,shawsilvermanslansky}.
(Supersymmetry avoids this problem by only having squarks couple to
gauginos with either pure L or R coupling, never mixing the two to
form a mass term.)  Here we examine without prejudice the
phenomenology assuming the same anomalous chromomagnetic moment for
each quark.  Separate analyses have been made for only the $t$
quark\cite{rizzocs,cheung,haberl} or also the $b$ quark\cite{silverman} having
the moment.  Formulas for the cross sections in high tranverse energy
jets\cite{silverman} and high transverse energy prompt photon
production\cite{cheungsil} have been given.

In this short contribution, we define a statisitical criteria for
comparing the sensitivity of new accelerators in energy and luminosity
to set limits on $\kappa' \equiv 1/\Lambda$.  $\Lambda$ is not to be
taken literally as the scale of the new phenomena, due to the complex
relation cited above.

\section{Simple Criteria for Statistical Sensitivity in High Transverse
Energy Jets and in Prompt Photon Production} 

Without a full Monte Carlo of the detector including energy
determination errors, we will treat here only the statistical
sensitivity of the various experiments.  Our criteria\cite{hinchliffe}
is to take bins of appropriate size for the energy range being
examined, and find the $E_T$ called $E_T^*$ at which the QCD cross
section statistical error bars are 10\%.  These will be bins in which
there are 100 QCD events.  We then explore the cross section due to
QCD plus the anomalous chromomagnetic moment contribution, and find
the value of $\kappa' \equiv 1/\Lambda$ or $\Lambda$ where the excess
over QCD is 10\% at this $E_T^*$.  These $E_T^*$ and $\Lambda$ are
shown in Table I.  Since the cross section is steeply falling, varying
the bin size by a factor of two makes only a small change in the value
of $E_T^*$ or $\Lambda$.  The limits in $|\eta|$ used are 0.9
for CDF and the Tevatron, and 1.0 for LHC.

We see from the table that $\Lambda$ sensitivity is roughly the same 
scale as the beam energy.  We also see that large increases in
luminosity do not increase $\Lambda$ proportionately even to the square
root of the luminosity.

\section{Equivalent Challenges in Theory and Systematical Errors}

To match a 10\% statistical uncertainty, the theory and systematic errors
must be reduced to the same amount.  Since structure functions
enter as a product of two of them, the dominant regions have to have
errors less than 5\% each.  The value of $\alpha_s$ at these transverse
energies must also be known better than 5\%.  The main systematic 
error is non-linearities in the energy measurement at these high $E_T$.
$d\sigma/dE_T$ falls at least as fast as $E_T^{-3}$ on dimensional
grounds, and also picks up some of the $(1-x_T)^n$ powers from the
structure functions, from $x_1 \approx x_2 \approx x_T 
\equiv E_T/(\sqrt{s}/2)$.  Using the minimum falloff of $E_T^{-3}$, 
a 3\% error on the linearity of $E_T$ at $E_T^*$ becomes a 10\% error
on the cross section.  

\begin{table*}[t!]
\leavevmode
\begin{center}
\caption{Table of High $E_T$ Bins at 10\% Statistical Error and
1-$\sigma$ Sensitivity for $\Lambda$ in That Bin}
\label{table1}
\vskip0.2in
\begin{tabular}{|l||r|r|r||r|r||r|r|}
\hline
 & & Integrated & &\multicolumn{2}{c|| }{$E_T$ Jets}  
&\multicolumn{2}{c|}{Photons} \\
\cline{5-8} \cline{5-8}
Accelerator & $E_{\rm cm}$ & Luminosity & Bin Width
& $E_T^*$ & $\Lambda$ & $E_T^*$ & $\Lambda$ \\
\hline
 & TeV & fb$^{-1}$ & GeV & GeV & TeV & GeV & TeV \\
\hline
Tevatron:& & & & & & & \\
         Run I  & 1.8 & 0.1 & 10 & 360 & 1.8 & 140 & 0.7 \\
         Run II & 2.0 & 2   & 20 & 490 & 2.8 & 260 & 1.5 \\
        Stretch & 2.0 & 10  & 20 & 540 & 3.3 & 325 & 1.9 \\
        TeV33   & 2.0 & 30  & 20 & 575 & 3.5 & 370 & 2.1 \\
\hline
        LHC     & 14  & 10  & 100 & 2500 & 13 & 1000 & 4.5 \\
        LHC     & 14  & 100 & 100 & 3100 & 17 & 1400 & 6.3 \\
\hline
\end{tabular}
\end{center}
\end{table*} 

These give goals for theory and energy measurement to be strived for
to make use of the high energy and luminosity achievable at the
Tevatron and at the LHC.  Finally, the statistical significance of
several bins in a row with deviation in the same direction can easily
be increased above the 1-$\sigma$ deviation of a single bin used here
by grouping all such bins into a large bin.  The details of doing this
in a specific case will depend on the other errors as well.

The authors acknowledge useful discussions with T.G. Rizzo, R. Peccei, R.
Harris, J. Whitmore, J. Soderquist and I. Hinchliffe.

\end{document}